\documentclass[prl,aps,showpacs,twocolumn]{revtex4}

\usepackage[dvips]{graphicx}
\usepackage{textcomp}

\begin{document}


\title {On the phases of Polonium}
\author{Matthieu J. Verstraete$^{1,2}$}
\email{matthieu.jean.verstraete@gmail.com}
\affiliation{$^{1}$
Nano-Bio group \& ETSF, Dpto. F\'{\i}sica de Mat., Univ. Pa\'{\i}s Vasco UPV/EHU, Centro F\'{\i}sica de Mat. CSIC-UPV/EHU and DIPC, E-20018 San Sebasti\'an, Spain}
\affiliation{$^{2}$ Unit\'e PCPM, Universit\'e Catholique de Louvain, Louvain-la-Neuve, Belgium}


\begin{abstract}
The thermodynamical properties of the main phases of metallic polonium are examined using
Density Functional Theory. The exceptional nature of the solid-solid phase transition of
$\alpha$ to $\beta$ Po is underlined: it induces a lowering in symmetry, from cubic to
rhombohedral, with increasing temperature. This is explained as the result of a delicate balance
between relativistic and entropic effects. Overall agreement with existing experimental data
is very good by state-of-the-art standards. The phonons of Po present Kohn
anomalies, and it is shown that the effect of spin-orbit interactions is the inverse of that in
normal metals: due to the non-spherical nature of the Fermi Surface, spin-orbit effects
reduce nesting and harden most phonon frequencies.
\end{abstract}

\maketitle

A frantic race is underway to find exotic new phases of elemental solids under
pressure\cite{oganov_2009_boron_high_pressure_short,
matsuoka_2009_lithium_semiconducting_phase, ma_2009_sodium_dielectric_phase_short,
mao_2003_superhard_graphite_phase_short}. With the advent of extreme diamond-anvil
cell experiments with pressures beyond 1 MBar, a whole slew
of new phases can be synthesized: with increasing pressure, the distance
between atoms decreases, electron hybridization changes, and the nature of
chemical bonds as well, changing coordination numbers and bonding
directionality, orbital occupations, and so on. This is particularly dramatic
in the recent demonstration that simple metals can become semiconducting if
valence and core orbitals are forced to hybridize, through the application of 
pressure\cite{matsuoka_2009_lithium_semiconducting_phase,
ma_2009_sodium_dielectric_phase_short}. 

Comparatively much rarer are purely \emph{temperature} induced solid to solid
phase transitions. Although about half\cite{tonkov_1992_phase_diagrams} of the
elements present some form of solid-solid transition with T, most are in very
limited pressure ranges, the critical pressure being more or less
constant with T. Almost all of these transitions either increase or
maintain the level of crystal symmetry (usually hexagonal to cubic, or cubic to
cubic). This is intuitive, as temperature will increase inter-atomic distances,
and tend to rotationally average orbital configurations, leading to high-symmetry
(most often BCC) then liquid structures. The exceptions are the usual suspects:
elements with complex bonding (boron, sulfur, phosphorous and bismuth),
and a few of those with delicate electronic shell effects, such as Ce or Dy. In
these systems, an increase in temperature can give rise to a solid phase with
\emph{lower} symmetry.

Polonium also falls into this category. It is one of the strangest elements of
the periodic table, and the only element which adopts the simple cubic
(SC)\cite{beamer_1949_Po_crystal_structure} structure at ambient pressure
($\alpha$-Po, joined by a few others such as Ca-III and
As-II\cite{mcmahon_2006_high_pressure_elements_review} at higher pressure).
Upon increasing temperature, between 290 and 330
K\cite{maxwell_1949_Po_thermodynamics}, Po transforms to the $\beta$ phase,
with a rhombohedral structure.

Despite its intriguing nature, experiments on Po are very difficult and rare,
and little is known about its properties. Only a few laboratories in the world
can synthesize Po in sufficient quantities to perform characterization. Due
to their high radioactivity (with a half life of at most 100 years, depending
on the isotope), samples heat and decay very quickly into Pb or Bi. 

By chemical analogy, Po should behave like Se and Te, its 4p and 5p cousins,
and adopt a trigonal spiral structure at low T. However, it was shown
recently\cite{legut_2007_Po_SC_relativity,
min_2006_Po_SC_structure_SO_Peierls_short} that the SC structure is stabilized
over the spiral structure by relativistic effects (though there is debate over
whether the scalar relativistic (SR) or spin-orbit (SO) terms dominate - see
comments to Ref.~\onlinecite{legut_2007_Po_SC_relativity}). It was predicted
that a simple rhombohedral structure would be stabilized under pressure. A
qualitative explanation for the appearance of the SC structure is that the
valence p states do not hybridize with s, because of strong relativistic
repulsion of the s\cite{kraig_2003_Po_structural_properties,
karen_2008_Po_relativistic_comment}. The p states can therefore sustain purely
right-angle bonding and an SC structure. Something similar probably happens in
Ca and As, due to the relative (but not necessarily relativistic) changes in
the s and p energy levels under pressure.

In this Letter, the vibrational and thermodynamical properties of crystalline
Po are calculated from first principles. First, the effect of the spin-orbit
interaction on the phonons of $\alpha$-Po is shown to be counter-intuitive,
hardening phonons and reducing the nesting of the Fermi surface (FS), the
opposite behavior of simple metals. Thermodynamical quantities are calculated
for the low temperature phase, and agreement with experiment confirms the
precision of the first principles methods. Finally, the transition temperature
of Polonium is calculated at 0 pressure, comparing the Gibbs free energy of the
different phases. Entropic and relativistic effects are delicately balanced and
determine the stability of the $\alpha$ and $\beta$ phases as a function of
temperature.

Density Functional Theory (DFT)\cite{martin_2004_dft_book}
and the perturbation theory built on top of it
(DFPT)\cite{baroni_2001_phonon_review} provide highly accurate calculations of
the total energy and vibrational properties of materials from first principles.
The implementation in the ABINIT package\cite{ABINIT_zkrist_short} is employed, with a
plane wave basis (cut off at 435 eV) and pseudopotentials in the local density
approximation\cite{kohn_1965_DFT_LDA}. Relativistic corrections are included through a pseudopotential
of the HGH form\cite{hartwigsen_1998_psp_hgh}. The corrections are decomposed
into SO and SR contributions. The calculations confirm the interpretation of
Ref.~\onlinecite{legut_2007_Po_SC_relativity} that the SR terms are sufficient
to stabilize the simple cubic phase. Calculations without the SR relativistic
terms will not be considered, because the of the structure of the
pseudopotentials used, and because for heavy elements this is quite
unrealistic. It is the spin-orbit term which has the largest influence on the
shape of the Fermi surface (which is the central quantity affecting phonon
structure). The DFPT formalism with SO is described in
Refs~\onlinecite{dalcorso_2007_SO_USPP_theory,
verstraete_2008_Pb_phonons_short}. The thermodynamics of polonium are treated
in the quasiharmonic approximation, fitting volume and temperature dependencies
as described by Grabowski et al.\cite{grabowski_2007_thermodynamics_fcc_metals}
The Brillouin Zone is sampled with 16$^3$ k-points for electrons and 8$^3$
q-points for phonons.

\begin{figure}[t] 
\includegraphics[clip=,width=0.95\linewidth]{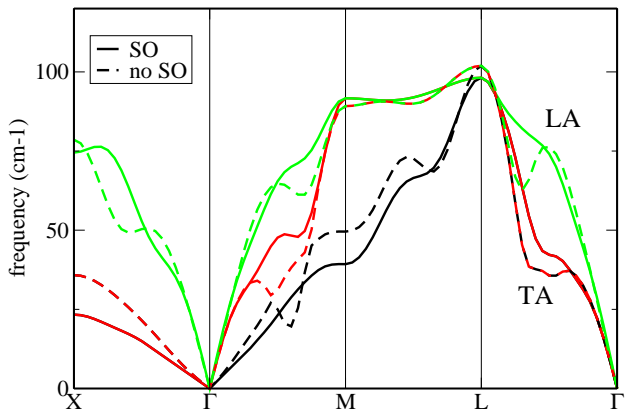}
\caption{The phonon band structure of SC polonium along high symmetry
directions of the Brillouin Zone, including the spin-orbit interaction (solid
line), and without it (dashed). Contrary to most normal metals, in Polonium
spin-orbit decreases nesting, removes Kohn anomalies, and hardens most phonon
frequencies.}
\label{sonoso_bst_fig}
\end{figure}

\begin{figure}[t] 
\includegraphics[clip=,width=0.47\linewidth]{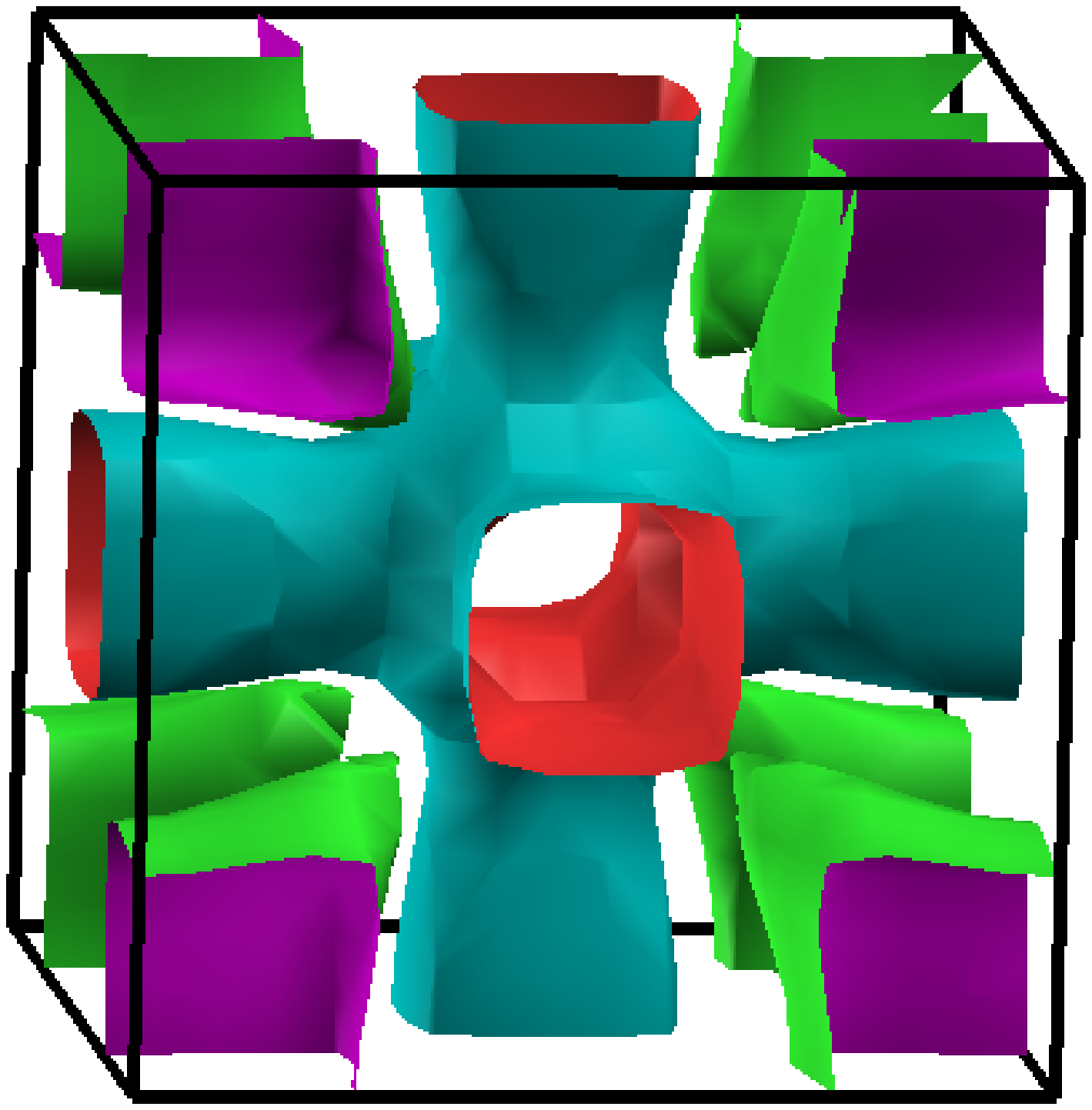}
\includegraphics[clip=,width=0.46\linewidth]{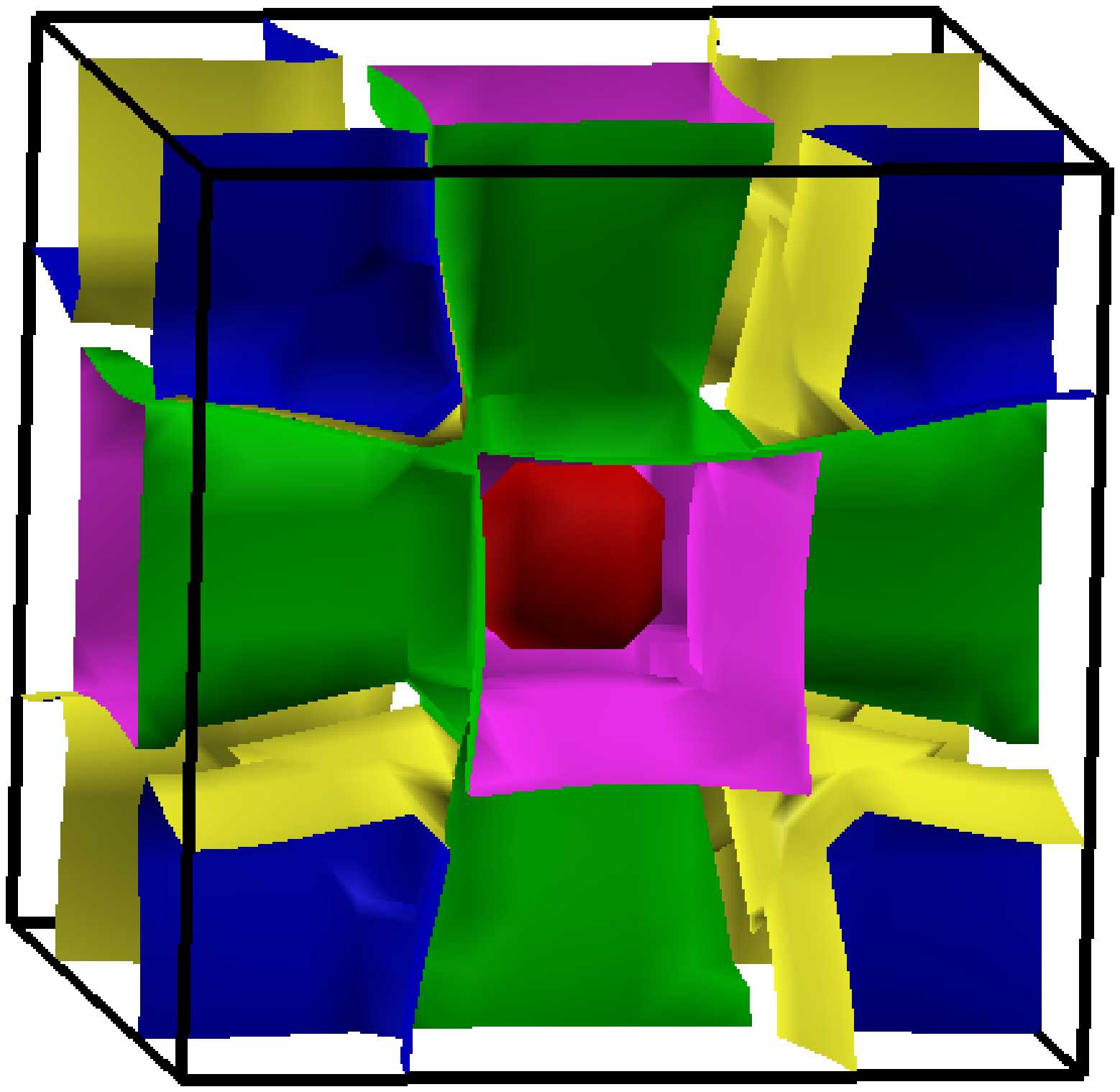}
\caption{Fermi surface of simple cubic Po, with (left) and without (right) the
spin-orbit interaction. The SO interaction reduces the nesting of the Fermi surface,
which in turn reduces the Kohn anomalies in Fig.~\ref{sonoso_bst_fig}.}
\label{sonoso_FS_fig}
\end{figure}

The $\alpha$ phase of polonium is simple cubic, with a lattice constant of
3.345~\AA{}. The theoretical equilibrium lattice constant is 3.335~\AA{}, in
good agreement, and will be used in the following. The phonon band structure of
Po is shown in Fig.~\ref{sonoso_bst_fig}, with and without the effect of the SO
interaction. In a simple metal, the Fermi surface departs only slightly from a
sphere. Kohn anomalies (KA)\cite{kohn_1959_anomaly} appear for phonons with
reciprocal space vectors $\vec{q}$ which are nested in the surface, i.e. those
$\vec{q}$ which connect many pairs of points on the surface. These phonons are
screened particularly strongly by the electrons, which renormalizes their
frequency. Distorting a spherical FS, in particular because of the SO
interaction, will always increase the nesting for certain $\vec{q}$ (sometimes
dramatically as in Pb\cite{verstraete_2008_Pb_phonons_short}). In polonium, the
opposite happens: due to the orbital nature and SC crystal structure, the Fermi
surface is more cubic than spherical (see Fig.~\ref{sonoso_FS_fig}). As a
result, the deformations due to the SO interaction will \emph{decrease} the
nesting, instead of increasing it. The small spherical hole pocket around
$\Gamma$ is also removed when the SO interaction is
added. These types of effects had already been seen in other response properties in
Ref.~\cite{min_2006_Po_SC_structure_SO_Peierls_short}. The result is quite
clear in the phonon bands of Fig.~\ref{sonoso_bst_fig}: when including the SO
interaction (going from dashed to full lines), the dips due to Kohn anomalies
are reduced, and the longitudinal (LA) frequency increases. The transverse (TA)
frequencies increase or decrease in different regions of the Brillouin Zone,
but also present KA which are reduced by the spin-orbit interaction.
From the vibrational calculations for the $\alpha$ phase with and without SO,
it is clear that inclusion of full relativistic effects is crucial to the
correct representation of the thermal properties of Po. On this basis, the
thermodynamical behavior of $\alpha$ and $\beta$ Po (and their relative
stability) will be examined in the following including both SR and SO terms.

\begin{figure}[t] 
\includegraphics[clip=,width=0.95\linewidth]{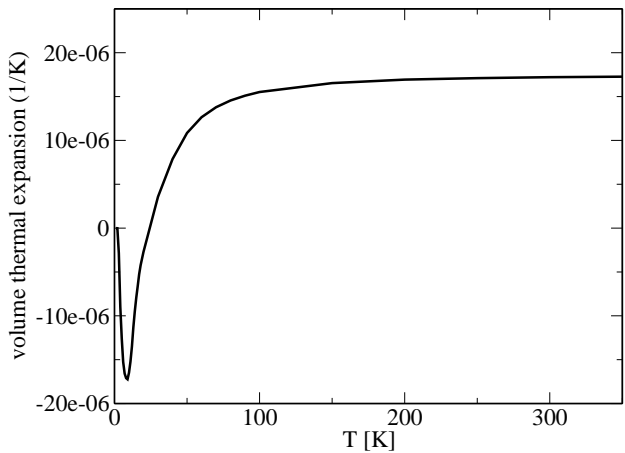}
\caption{Thermal expansion coefficient of Po as a function of temperature.
Experiments\cite{brocklehurst_1957_Po_thermal_expansion} give 23 $\pm 1.5$ 10$^{-6}$
$K^{-1}$ at 300 Kelvin.}
\label{th_expan_fig}
\end{figure}

Thermodynamical quantities are calculated from the phonon bands of the bulk
metal. To capture effects of thermal expansion, full phonon calculations are
also performed for a volume 3\% larger than equilibrium, and the mode-dependent
Gr\"uneisen parameters are extracted (see e.g.
Ref.~\onlinecite{grabowski_2007_thermodynamics_fcc_metals} for standard
formulae). The thermal expansion of Po is shown in Fig.~\ref{th_expan_fig} as a
function of temperature; it is initially negative, which is
found in many materials. For low temperatures the mode behavior near the
Brillouin Zone center is crucial. The contribution from the smallest $\vec{q}$
vectors (up to 1/8 of Brillouin Zone) is therefore fit to explicit calculations
at small $\vec{q}$ and integrated numerically. 

The principal thermodynamical quantities for which there is experimental data
are the thermal expansion coefficient and the heat capacity. Brocklehurst et
al.\cite{brocklehurst_1957_Po_thermal_expansion} found a thermal expansion
coefficient of $23 \cdot 10^{-6}$ $\pm$ $1.5 \cdot 10^{-6}$ K$^{-1}$ at 298 K.
First principles calculations give $16 \cdot 10^{-6}$ K$^{-1}$ which is in
reasonable agreement; the residual difference is probably due to anharmonic
effects.
The tabulated heat capacity of polonium at 300~K is
C$_P$ = 26.4 J mol$^{-1}$ K$^{-1}$\cite{dean_1999_lange_handbook_of_chemistry},
quite close to the ab initio value of 24.9 J mol$^{-1}$ K$^{-1}$ found here. 
The correction due to SO coupling is slight ($+0.1$ J mol$^{-1}$ K$^{-1}$).
A fit to Murnaghan's\cite{murnaghan_1944_equation_of_state} equation of state for $\alpha$-Po
gives a bulk modulus $B=39.45$~GPa, volume 250~bohr$^3$ and $B'=4.89$. 



\begin{figure}[b] 
\includegraphics[clip=,width=0.95\linewidth]{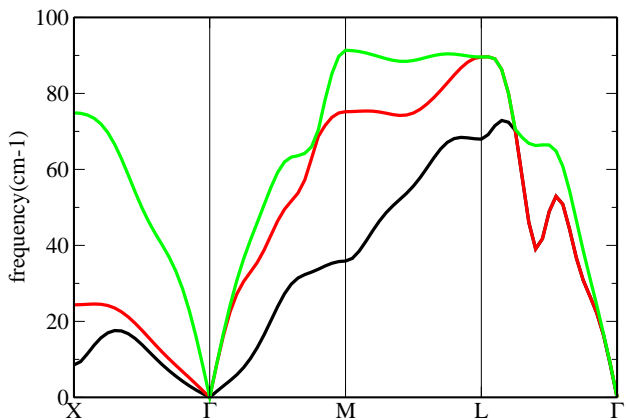}
\caption{The phonon band structure of $\beta$ polonium along high symmetry
directions of the Brillouin Zone, including the spin-orbit interaction}
\label{beta_so_bst_fig}
\end{figure}

We now turn to the thermodynamics of $\beta$-Po, and the explanation of why
this lower symmetry phase is favored at higher temperatures, examining the
contributions to the free energy which stabilize $\alpha$ or $\beta$ Po at
different temperatures. The beta phase of Polonium is rhombohedral, with a
lattice constant of 3.359~\AA{} and an angle of $98^{\circ}$ 13'. It
corresponds to a cube which has been slightly compressed along the (111)
diagonal, and possesses fairly low symmetry. As noted in previous studies, the
structure is intermediate between the SC and a body-centered-cubic phase, which
is metastable at the equilibrium volume. For an angle of $60^{\circ}$, a
face-centered-cubic phase is also metastable. For compressed volumes,
calculations confirm the results of
Ref.~\onlinecite{legut_2007_Po_SC_relativity} that a single rhombohedral phase
becomes stable for volumes smaller than about $0.93 V_0$ (for a calculated
pressure of 1 GPa), with an angle larger than $90^{\circ}$. Then, at
compressions below 90\% (for a calculated pressure between 1.5 and 2 GPa), a
second phase is also stabilized, with an angle smaller than $90^{\circ}$ (as
found in Ref.~\onlinecite{legut_2007_Po_SC_relativity}). Fitting an equation
of state to the $E(V)$ relation gives an analytical form for the basic
variation of the energy with volume. This is not simple with $\beta$-Po, for
two reasons. First, phases with the equilibrium volume and unit cell angles
between $90^{\circ}$ and $100^{\circ}$ are never even metastable with respect
to $\alpha$-Po. Second, fitting the equation of state to compressed
rhombohedral phases is not useful, as an unrealistically large equilibrium
volume is found. Given the good performance of the DFT methods for the SC
phase, the experimental unit cell is used to compute the phonon bands of
$\beta$-Po, and the equation of state is fit for a constant rhombohedral angle
(giving $B=40.15$ GPa, volume 247~bohr$^3$ and $B'=4.49$). Thermal
expansion effects are then taken into account as for the $\alpha$ phase. This
will be a very good approximation for volumes and temperatures close to our
reference point, which is the experimental volume at around
348~K\cite{beamer_1949_Po_crystal_structure}. It has been checked that
unstable modes do not appear in the various compressed or distorted structures.

\begin{figure}[t] 
\includegraphics[clip=,width=0.95\linewidth]{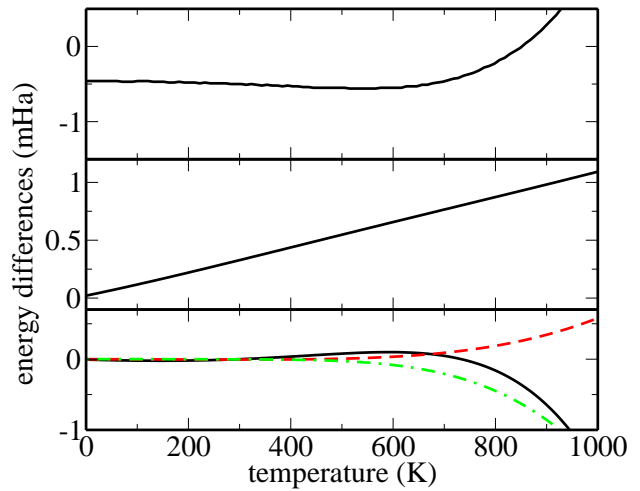}
\caption{Free energy differences (mHa) between $\beta$ and $\alpha$ phases of
Polonium, as a function of temperature (Kelvin). Positive values favor the
stability of $\beta$ Polonium. Top panel: total energy at 0 K and thermally
expanded volume. Middle panel: vibrational free energy at constant volume
$F_{vib}$. Bottom panel: volume-correction $\Delta F_{vib}$ (solid),
electronic free energy at constant volume $F_{el}$ (dash), and
volume-correction $\Delta F_{el}$ (dot-dash). Differences are small below 600
K. It is the two vibrational free energy terms that determine the transition
temperature from $\alpha$ to $\beta$ Po.}
\label{free_en_terms_fig}
\end{figure}

The phonon band structure of $\beta$-Po is shown in Fig.~\ref{beta_so_bst_fig},
along the same high-symmetry directions as in Fig.~\ref{sonoso_bst_fig}
(the SC nomenclature is used for simplicity).
The phonons are calculated at 0 K (thermal expansion is later taken into account as for
the $\alpha$ phase), and would change somewhat due to anharmonic effects (see
e.g. Ref.~\onlinecite{souvatzis_2008_finitet_phonons}). Work in this direction is
ongoing, but the current agreement with experiment suggests that anharmonic
effects are small. It has been checked that lowering the electronic
temperature does not lead to unstable phonons (in particular for the dip between
$L$ and $\Gamma$), such that the thermodynamics described by the 0 K phonons is
physical. The free energy is calculated for $\alpha$ and $\beta$-Po, adding
the ground state energy to the vibrational and
electronic free energies. The free energies are calculated at a second volume
$V'$, and an additional correction is added, taking the thermal expansion into
account (bottom panel of Fig.~\ref{free_en_terms_fig}): $\Delta F_x(T) =
(F_x(V',T)-F_x(V_0,T)) \cdot (V(T)-V_0)/(V'-V_0)$ where x is vibrational or
electronic. The volume $V(T)$ is obtained by inverting $P = -\partial F /
\partial V = 0$, and inserted in the free energy to give the Gibbs free energy
as a function of temperature. The differences in the free energy components 
between the $\alpha$ and the $\beta$ phases are shown in
Fig.~\ref{free_en_terms_fig}. In Ref.~\onlinecite{legut_2007_Po_SC_relativity}
the relativistic stabilization of $\alpha$-Po is calculated to be 0.15~mHa. Up to
$\sim$700 K, the total energy difference between the phases (including the
relativistic stabilization for both) is ca. -0.5~mHa. At 350~K the
vibrational entropy difference is of the order of 0.4~mHa, and the other terms
also favor the $\beta$ phase. The transition can thus be explained by the
vibrational entropy terms in $\beta$-Po overcoming the 0 K total energy
difference.

Comparing the full Gibbs free energy of $\alpha$-Po and $\beta$-Po
determines the transition temperature $T_c$. The theoretical
$T_c$ (at 0 pressure) from the $\alpha$ to the $\beta$ phase
is 370~$\pm 20$~K. The error is estimated from the convergence of the free
energy fit parameters, and does not take into account anharmonic corrections.
This is in quite good agreement with experimental data which give a
range of transition temperatures from 290 to 330~K. Omitting the thermal
expansion or any of the free energy terms removes the transition and stabilizes
$\alpha$-Po for all temperatures.

In summary, this Letter presents an ab initio study of the different
thermodynamical phases of crystalline Po. The effect of spin-orbit coupling on
the phonon band structure of simple cubic Po is first shown to be the opposite
of that on simple metals: as the initial Fermi surface (without SO) is strongly
nested, the SO correction reduces the nesting and the resulting Kohn anomalies,
and hardens most phonon frequencies. The inclusion of SO effects is essential
to the correct representation of the phonons. Due to the delicate bonding
characteristics of Po, whose low temperature simple cubic phase is stabilized
by relativistic effects, increasing the temperature has the completely
counter-intuitive effect of stabilizing a less symmetric phase (rhombohedral
$\beta$-Po). This transition originates in the competition of relativistic
stabilizing effects for the $\alpha$ phase and slightly more favorable
vibrational entropy and thermal expansion for the $\beta$ phase. A number of
other elements are likely to have temperature behaviors similar to Po: either
due to relativistic or pressure effects, complex bonding environments open the
way for radically different symmetry breaking as a function of temperature,
when entropic effects are of the same order of magnitude as the energy
differences between bonding types.

The author wishes to thank M. \v{S}ob, N. Helbig, and O. Eriksson for useful discussions.
This research has been supported by the Belgian FNRS, the EU FP6 project DNA
Nanodevices (G.A. 029192), and the EU FP7 through the ETSF I3 e-I3 project
(G.A. 211956). Computer time was provided by UCLouvain-CISM, CINES, and the Red
Espa\~nola de Supercomputacion.


\begin{thebibliography}{25}
\expandafter\ifx\csname natexlab\endcsname\relax\def\natexlab#1{#1}\fi
\expandafter\ifx\csname bibnamefont\endcsname\relax
  \def\bibnamefont#1{#1}\fi
\expandafter\ifx\csname bibfnamefont\endcsname\relax
  \def\bibfnamefont#1{#1}\fi
\expandafter\ifx\csname citenamefont\endcsname\relax
  \def\citenamefont#1{#1}\fi
\expandafter\ifx\csname url\endcsname\relax
  \def\url#1{\texttt{#1}}\fi
\expandafter\ifx\csname urlprefix\endcsname\relax\def\urlprefix{URL }\fi
\providecommand{\bibinfo}[2]{#2}
\providecommand{\eprint}[2][]{\url{#2}}

\bibitem[{\citenamefont{Oganov~{\it et
  al.}}(2009)}]{oganov_2009_boron_high_pressure_short}
\bibinfo{author}{\bibfnamefont{A.~R.} \bibnamefont{Oganov~{\it et al.}}},
  \bibinfo{journal}{Nature} \textbf{\bibinfo{volume}{457}},
  \bibinfo{pages}{863} (\bibinfo{year}{2009}).

\bibitem[{\citenamefont{Matsuoka and
  Shimizu}(2009)}]{matsuoka_2009_lithium_semiconducting_phase}
\bibinfo{author}{\bibfnamefont{T.}~\bibnamefont{Matsuoka}} \bibnamefont{and}
  \bibinfo{author}{\bibfnamefont{K.}~\bibnamefont{Shimizu}},
  \bibinfo{journal}{Nature} \textbf{\bibinfo{volume}{458}},
  \bibinfo{pages}{186} (\bibinfo{year}{2009}).

\bibitem[{\citenamefont{Ma~{\it et
  al.}}(2009)}]{ma_2009_sodium_dielectric_phase_short}
\bibinfo{author}{\bibfnamefont{Y.}~\bibnamefont{Ma~{\it et al.}}},
  \bibinfo{journal}{Nature} \textbf{\bibinfo{volume}{458}},
  \bibinfo{pages}{182} (\bibinfo{year}{2009}).

\bibitem[{\citenamefont{Mao~{\it et
  al.}}(2003)}]{mao_2003_superhard_graphite_phase_short}
\bibinfo{author}{\bibfnamefont{W.~L.} \bibnamefont{Mao~{\it et al.}}},
  \bibinfo{journal}{Science} \textbf{\bibinfo{volume}{302}},
  \bibinfo{pages}{425} (\bibinfo{year}{2003}).

\bibitem[{\citenamefont{Tonkov}(1992)}]{tonkov_1992_phase_diagrams}
\bibinfo{author}{\bibfnamefont{E.~Y.} \bibnamefont{Tonkov}},
  \emph{\bibinfo{title}{High Pressure phase transformations}}
  (\bibinfo{publisher}{Gordon and Breach}, \bibinfo{address}{Philadelphia},
  \bibinfo{year}{1992}).

\bibitem[{\citenamefont{Beamer and
  Maxwell}(1949)}]{beamer_1949_Po_crystal_structure}
\bibinfo{author}{\bibfnamefont{W.~H.} \bibnamefont{Beamer}} \bibnamefont{and}
  \bibinfo{author}{\bibfnamefont{C.~E.} \bibnamefont{Maxwell}},
  \bibinfo{journal}{J Chem Phys} \textbf{\bibinfo{volume}{17}},
  \bibinfo{pages}{1293} (\bibinfo{year}{1949}).

\bibitem[{\citenamefont{McMahon and
  Nelmes}(2006)}]{mcmahon_2006_high_pressure_elements_review}
\bibinfo{author}{\bibfnamefont{M.~I.} \bibnamefont{McMahon}} \bibnamefont{and}
  \bibinfo{author}{\bibfnamefont{R.~J.} \bibnamefont{Nelmes}},
  \bibinfo{journal}{Chem Soc Rev} \textbf{\bibinfo{volume}{35}},
  \bibinfo{pages}{943–963} (\bibinfo{year}{2006}).

\bibitem[{\citenamefont{Maxwell}(1949)}]{maxwell_1949_Po_thermodynamics}
\bibinfo{author}{\bibfnamefont{C.~E.} \bibnamefont{Maxwell}},
  \bibinfo{journal}{J Chem Phys} \textbf{\bibinfo{volume}{17}},
  \bibinfo{pages}{1288} (\bibinfo{year}{1949}).

\bibitem[{\citenamefont{Legut et~al.}(2007)\citenamefont{Legut, Fri\'{a}k, and
  \v{S}ob}}]{legut_2007_Po_SC_relativity}
\bibinfo{author}{\bibfnamefont{D.}~\bibnamefont{Legut}},
  \bibinfo{author}{\bibfnamefont{M.}~\bibnamefont{Fri\'{a}k}},
  \bibnamefont{and} \bibinfo{author}{\bibfnamefont{M.}~\bibnamefont{\v{S}ob}},
  \bibinfo{journal}{Phys Rev Lett} \textbf{\bibinfo{volume}{99}},
  \bibinfo{pages}{016402} (\bibinfo{year}{2007}).

\bibitem[{\citenamefont{{Min {\it et
  al.}}}(2006)}]{min_2006_Po_SC_structure_SO_Peierls_short}
\bibinfo{author}{\bibfnamefont{B.~I.} \bibnamefont{{Min {\it et al.}}}},
  \bibinfo{journal}{Phys Rev B} \textbf{\bibinfo{volume}{73}},
  \bibinfo{pages}{132102} (\bibinfo{year}{2006}).

\bibitem[{\citenamefont{Kraig et~al.}(2004)\citenamefont{Kraig, Roundy, and
  Cohen}}]{kraig_2003_Po_structural_properties}
\bibinfo{author}{\bibfnamefont{R.~E.} \bibnamefont{Kraig}},
  \bibinfo{author}{\bibfnamefont{D.}~\bibnamefont{Roundy}}, \bibnamefont{and}
  \bibinfo{author}{\bibfnamefont{M.~L.} \bibnamefont{Cohen}},
  \bibinfo{journal}{Sol. State Comm.} \textbf{\bibinfo{volume}{129}},
  \bibinfo{pages}{411–413} (\bibinfo{year}{2004}).

\bibitem[{\citenamefont{Karen}(2008)}]{karen_2008_Po_relativistic_comment}
\bibinfo{author}{\bibfnamefont{P.}~\bibnamefont{Karen}},
  \bibinfo{journal}{Physics Today} \textbf{\bibinfo{volume}{September}},
  \bibinfo{pages}{10} (\bibinfo{year}{2008}).

\bibitem[{\citenamefont{Martin}(2004)}]{martin_2004_dft_book}
\bibinfo{author}{\bibfnamefont{R.~M.} \bibnamefont{Martin}},
  \emph{\bibinfo{title}{Electronic Structure: Basic Theory and Practical
  Methods}} (\bibinfo{publisher}{Cambridge University Press},
  \bibinfo{address}{Cambridge}, \bibinfo{year}{2004}).

\bibitem[{\citenamefont{Baroni et~al.}(2001)\citenamefont{Baroni, de~Gironcoli,
  Dal~Corso, and Giannozzi}}]{baroni_2001_phonon_review}
\bibinfo{author}{\bibfnamefont{S.}~\bibnamefont{Baroni}},
  \bibinfo{author}{\bibfnamefont{S.}~\bibnamefont{de~Gironcoli}},
  \bibinfo{author}{\bibfnamefont{A.}~\bibnamefont{Dal~Corso}},
  \bibnamefont{and}
  \bibinfo{author}{\bibfnamefont{P.}~\bibnamefont{Giannozzi}},
  \bibinfo{journal}{Rev. Mod. Phys.} \textbf{\bibinfo{volume}{73}},
  \bibinfo{pages}{515} (\bibinfo{year}{2001}).

\bibitem[{\citenamefont{{Gonze {\it et al.}}}(2005)}]{ABINIT_zkrist_short}
\bibinfo{author}{\bibfnamefont{X.}~\bibnamefont{{Gonze {\it et al.}}}},
  \bibinfo{journal}{Zeit. Krist.} \textbf{\bibinfo{volume}{220}},
  \bibinfo{pages}{558} (\bibinfo{year}{2005}).

\bibitem[{\citenamefont{Kohn and Sham}(1965)}]{kohn_1965_DFT_LDA}
\bibinfo{author}{\bibfnamefont{W.}~\bibnamefont{Kohn}} \bibnamefont{and}
  \bibinfo{author}{\bibfnamefont{L.~J.} \bibnamefont{Sham}},
  \bibinfo{journal}{Phys. Rev.} \textbf{\bibinfo{volume}{140}},
  \bibinfo{pages}{A 1133 } (\bibinfo{year}{1965}).

\bibitem[{\citenamefont{Hartwigsen et~al.}(1998)\citenamefont{Hartwigsen,
  Goedecker, and Hutter}}]{hartwigsen_1998_psp_hgh}
\bibinfo{author}{\bibfnamefont{C.}~\bibnamefont{Hartwigsen}},
  \bibinfo{author}{\bibfnamefont{S.}~\bibnamefont{Goedecker}},
  \bibnamefont{and} \bibinfo{author}{\bibfnamefont{J.}~\bibnamefont{Hutter}},
  \bibinfo{journal}{Phys. Rev. B} \textbf{\bibinfo{volume}{58}},
  \bibinfo{pages}{3641} (\bibinfo{year}{1998}).

\bibitem[{\citenamefont{Dal~Corso}(2007)}]{dalcorso_2007_SO_USPP_theory}
\bibinfo{author}{\bibfnamefont{A.}~\bibnamefont{Dal~Corso}},
  \bibinfo{journal}{Phys. Rev. B} \textbf{\bibinfo{volume}{76}},
  \bibinfo{pages}{054308} (\bibinfo{year}{2007}).

\bibitem[{\citenamefont{{Verstraete {\it et
  al.}}}(2008)}]{verstraete_2008_Pb_phonons_short}
\bibinfo{author}{\bibfnamefont{M.~J.} \bibnamefont{{Verstraete {\it et al.}}}},
  \bibinfo{journal}{Phys Rev B} \textbf{\bibinfo{volume}{78}},
  \bibinfo{pages}{045119} (\bibinfo{year}{2008}).

\bibitem[{\citenamefont{Grabowski et~al.}(2007)\citenamefont{Grabowski, Hickel,
  and Neugebauer}}]{grabowski_2007_thermodynamics_fcc_metals}
\bibinfo{author}{\bibfnamefont{B.}~\bibnamefont{Grabowski}},
  \bibinfo{author}{\bibfnamefont{T.}~\bibnamefont{Hickel}}, \bibnamefont{and}
  \bibinfo{author}{\bibfnamefont{J.}~\bibnamefont{Neugebauer}},
  \bibinfo{journal}{Phys. Rev. B} \textbf{\bibinfo{volume}{76}},
  \bibinfo{pages}{024309} (\bibinfo{year}{2007}).

\bibitem[{\citenamefont{Kohn}(1959)}]{kohn_1959_anomaly}
\bibinfo{author}{\bibfnamefont{W.}~\bibnamefont{Kohn}}, \bibinfo{journal}{Phys.
  Rev. Lett.} \textbf{\bibinfo{volume}{2}}, \bibinfo{pages}{393}
  (\bibinfo{year}{1959}).

\bibitem[{\citenamefont{Brocklehurst et~al.}(1957)\citenamefont{Brocklehurst,
  Goode, and Vassamillet}}]{brocklehurst_1957_Po_thermal_expansion}
\bibinfo{author}{\bibfnamefont{R.}~\bibnamefont{Brocklehurst}},
  \bibinfo{author}{\bibfnamefont{J.}~\bibnamefont{Goode}}, \bibnamefont{and}
  \bibinfo{author}{\bibfnamefont{L.}~\bibnamefont{Vassamillet}},
  \bibinfo{journal}{J. Chem. Phys.} \textbf{\bibinfo{volume}{27}},
  \bibinfo{pages}{985} (\bibinfo{year}{1957}).

\bibitem[{\citenamefont{Dean}(1999)}]{dean_1999_lange_handbook_of_chemistry}
\bibinfo{editor}{\bibfnamefont{J.}~\bibnamefont{Dean}}, ed.,
  \emph{\bibinfo{title}{Lange's Handbook of Chemistry}}
  (\bibinfo{publisher}{McGraw-Hill}, \bibinfo{year}{1999}),
  \bibinfo{edition}{15th} ed.

\bibitem[{\citenamefont{Murnaghan}(1944)}]{murnaghan_1944_equation_of_state}
\bibinfo{author}{\bibfnamefont{F.}~\bibnamefont{Murnaghan}},
  \bibinfo{journal}{Proc. Nat. Acad. Sci.} \textbf{\bibinfo{volume}{7}},
  \bibinfo{pages}{244} (\bibinfo{year}{1944}).

\bibitem[{\citenamefont{Souvatzis et~al.}(2008)\citenamefont{Souvatzis,
  Eriksson, Katsnelson, and Rudin}}]{souvatzis_2008_finitet_phonons}
\bibinfo{author}{\bibfnamefont{P.}~\bibnamefont{Souvatzis}},
  \bibinfo{author}{\bibfnamefont{O.}~\bibnamefont{Eriksson}},
  \bibinfo{author}{\bibfnamefont{M.I.}~\bibnamefont{Katsnelson}},
  \bibnamefont{and} \bibinfo{author}{\bibfnamefont{S.P.}~\bibnamefont{Rudin}},
  \bibinfo{journal}{Phys. Rev. Lett.} \textbf{\bibinfo{volume}{100}},
  \bibinfo{pages}{095901} (\bibinfo{year}{2008}).

\end{thebibliography}

\end{document}